\documentclass{du-journals}
\journal{Iranian Journal of Astronomy and Astrophysics}
\title{Extracting the Temperature of a Coronal Loop in the Solar Active Region 11092}
\subtitle{Temperature of a Coronal Loop}
\author[1]{Narges Fathalian}
\address[1]{Department of Physics, Payame Noor University (PNU), P.O.Box 19395-3697, Tehran, Iran;\\narges.fathalian@gmail.com}
\begin{document}
\begin{abstract}
Extracting the temperature of coronal loops is effective in the analysis of solar active region's loops and helps in better understanding of coronal events. To this end, various methods have already been developed like the method developed by Aschwanden et al. 2015 \cite{AschwandenB2015} which is based on Gaussian fit for Differential Emission Measure (DEM). Here, we use the intensity ratios of the images in three different wavelengths and the temperature response functions of AIA to extract the temperature of the loop.
In this paper, we use EUV images of the solar active region 11092 taken by AIA instrument of the SDO satellite at 171, 193, and 211 $A^{0}$ wavelengths, at 1th of August, 2010. We select a loop in a subregion of 11092 and extract its temperature by the help of intensity profiles in different wavelengths and thermal response functions of different filters. In this subregion cooling of the loop happens and in the selected loop, highest relative intensity of the wavelengths of 171 $A^{0}$ to 193 $A^{0}$ was obtained to be $0.76$ and this number was estimated to correspond to the temperature of 1.3 million degree of Kelvin, which is the maximum temperature point of this loop's internal area. The highest values of the intensity ratios at the wavelengths of 211$A^{0}$ to 193$A^{0}$, and 211$A^{0}$ to 171$A^{0}$ are $0.22$ and $0.25$, which correspond to temperature values around 10M and 1.4M Kelvin in sequence, related to the temperature of hotter and more superficial points of the loop respectively. These values are very sensitive to time and differ in time series of this event of the loop intensity variation.
\end{abstract}


\section{Introduction}

Solar images in the ranges of Extreme ultraviolet (EUV), Soft X-Ray (SXR) and other ranges of wavelengths are recorded by orbiting telescopes, SoHO, TRACE, STEREO, Hinode and other spatial missions. The last satellite, which transmits the uttermost volume of these images, and its data is available, is the Solar Dynamic Observatory (SDO) which launched on its orbit in 11th of February, 2010. The SDO satellite sends us about $1.6$ terabytes of data per day. EUV solar images give us a lot of information about the structure of the chromosphere, the transition zone, the corona, and the sun's magnetic field. In the analysis of the corona active region loops, extracting thermal details of coronal loops is of significant importance.

Several different methods have been developed to investigate thermal structure of the coronal loops and strands. The thermal stability of the coronal loops has been the subject of research done by Habbal and Rosner, 1979 \cite{Habbal1979} (and references cited therein). MacClymont and Craig (1984) \cite{MacClymont1984} also discussed in the field of thermal stability that a symmetric coronal temperature perturbation must be assisted by a pressure fluctuation. They concluded that, for the case of uniform heating, coronal loops are impartially stable. Schmelz et al. 2010 \cite{Schmelz2010} analyzed a coronal loop with AIA, observed on  Aug 3rd 2010, by some differential emission measure (DEM) curves, claiming a multithermal rather than an isothermal DEM distribution (for cross-sectional temperature of the loop). However, Aschwanden and Boerner (2010)\cite{AschwandenB2010} criticized the method of background subtraction which Schmelz et al. used. They claimed that it was their background subtraction method which causes their inferred result of a multithermal loop. This controversy of whether the cross-field temperatures of coronal loops are multithermal or isothermal, continued by Schmelz et al. 2013's work \cite{Schmelz2013} (similar to Schmelz et al. 2011 \cite{Schmelz2011}). They analyzed twelve loops to understand their cross-field temperature distributions and reveal the loop substructure. They found that warmer loops require broader DEMs. Afterwards, Schmelz et al. 2014\cite{Schmelz2014} found indications of a relationship between the DEM weighted temperature and the cross-field DEM width for coronal loops and discussed that cooler loops tend to have narrower DEM widths. This could imply that fewer strands are seen emitting in the later cooling phase, which they claim could potentially resolve the controversy.
Dependence of coronal loop temperature on loop length and the strength of magnetic field is also a favorite subject. For instance, Dahlburg et al. 2018\cite{Dahlburg2018} probed the temperature characteristics of solar coronal loops over a wide range of lengths and magnetic field strengths by means of numerical simulations and observed a very high correlation between magnetic field strength and maximum temperature. The effect of temperature inhomogeneity on the periods and the damping times of the standing slow modes in stratified solar coronal loops is also studied (see for example, Abedini et al., 2012\cite{Abedini2012}). In this field, Aschwanden et al. 2015 \cite{AschwandenB2015} (as well as 2013\cite{AschwandenB2013}) developed a method for extracting the loop temperature which is based on Gaussian fit for Differential Emission Measure, named spatially-synthesized Gaussian DEM forward-fitting method (DEM here after).
 Here, we develop a different method using the intensity ratios of the images in three different wavelengths and the temperature response functions of AIA to extract the temperature of the loop. The specifications of the selected Data are treated in Section 2. In section 3, the image analysis processes are explained. The background subtraction method is described in section 4 and the intensity profiles are shown there. Our thermal investigations are expressed in section 5. And in section 6, we explained our results and conclusions.

\section{Data}

SDO images are saving and categorizing by different instruments and in different filters. SDO has HMI (Helioseismic and Magnetic Imager), AIA (Atmospheric Imaging Assembly), and EVE (Extreme Ultraviolet Variability Experiment) instruments; in this investigation due to the mentioned subject, we use AIA instrument's data. We selected active region 11092, located at (-467, 208) arcsec. (The loop area is marked in Figure 1). After comparing several active regions based on their loops shapes, this smooth one, witch its loops have more definite shapes, was preferred. Of course, it is necessary to mention that there are also errors and missed images in the categorized data and some of the images are imperfect, and hence, before final selection and beginning the analysis of the selected region, we should make sure of the existence and safety of the data in the desired period of time. Appropriate time period, data existence in the categories, and nearness to the center of disk, not to the edges, were the other criterions for the selection of this region. In the first step, a four and half an hour data set (13:30 UT-18:00 UT) taken by the AIA on board the Solar Dynamics Observatory (Brown et al. 2010 \cite{Brown2010}, Lemen et al. 2012 \cite{Lemen2012}, Pesnell et al. 2012 \cite{Pesnell2012}) is used to find the desired event. This selected data set includes 550 images in each wavelength. These images are recorded in the date of 2010:08:01, and are captured in cadence of 12 seconds sequentially. The images size is 1000$\times$833 pixels and those are at the wavelengths of 171 $A^{0}$(Fe IX ion, quiet corona, upper transition region), 211$A^{0}$(Fe XIV ion, active region corona), and 193$A^{0}$(Fe XXIV ion and Fe XII, corona and hot flare plasma), in which the loops are visible more distinctively.

\begin{figure} %Fig.~1
 \centerline{\includegraphics[width=0.9\textwidth]{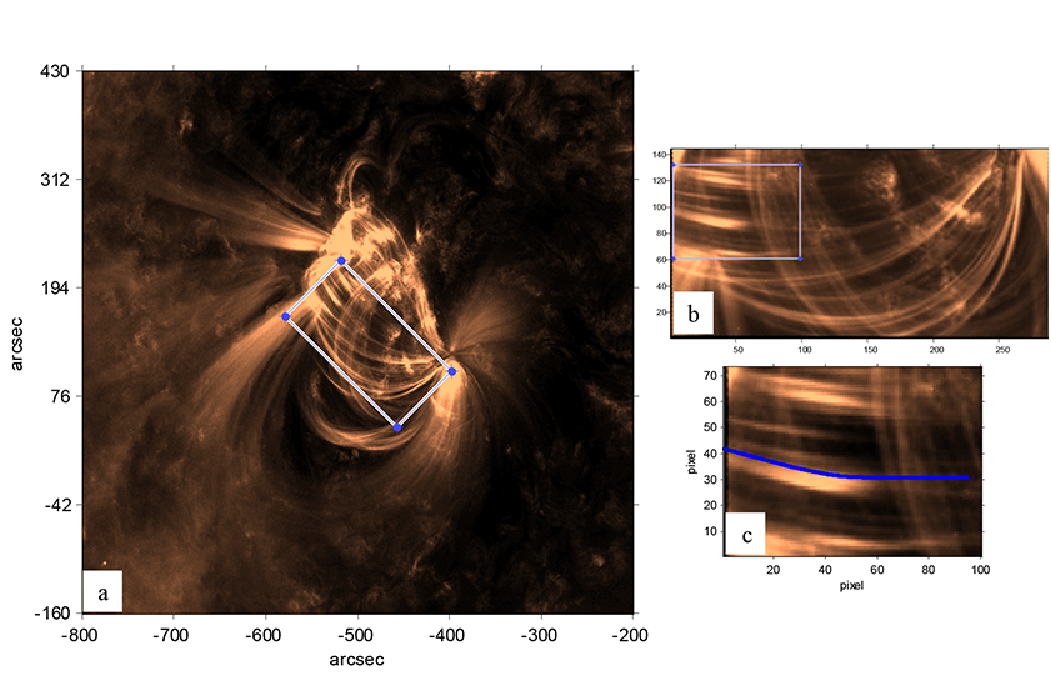}}
 \caption[]{a) The 11092 active region (located in (-467, 208) arcsec) and its around area, on 2010 August 1, at the wavelength of 174$A^0$. The selected subregion is marked by a box in the image. b) The selected subregion marked in Fig1.a is displayed here (in pixles). c) Final cut and the extracted loop marked in blue.}
 \label{fig1}
\end{figure}
\section{Image Analysis}

Receipted images are in fits format and we need special software packages to read and manipulate them. We also consider the effect of differential rotation of the sun. In the fits files headers, the index of CRPIX1 shows the reference pixel in the x-direction and by plotting this item, we could have a criterion of sun motion along time duration of data. We apply SSWIDL procedure, prep.pro code to preprocess the received data. Based on the favorite features of the corona, different preprocessing and extracting methods have been developed and used till now (e.g. Alipour et al. 2012 \cite{Alipour2012}, Tajfirouze and Safari 2012 \cite{Tajfirouze2010}, Alipour and Safari 2015 \cite{Alipour2015}, Yousefzadeh et al. 2015 \cite{Yousefzadeh2015}, Arish et al. 2016\cite{Arish2016}, Honarbakhsh et al. 2016 \cite{Honarbakhsh2016}).
One of the methods for primary study of the loops images is to view images in the form of time series or movie (by help of the xstepper code in the IDL software for instance). In addition, to follow an event in different wavelengths, it is appropriate to overlay the images in different wavelengths by assigning different colors to each one and to view the desired phenomena simultaneously in different wavelengths. Furthermore, to concentrate on interesting physical time variable events, it is suitable to investigate intensity difference of the images, which is obtained by subtracting each image from its previous one. After surveying active region of 11092, following above-mentioned procedure, flow is not observed in such a manner to oscillate during one loop. But complicated and different flows are observed moving along and across a bundle of loops and with temperature changes, and therefore ionizations changes, intensities in different parts of the loops are continually changing. Looking at images in the form of time series and then removing the corrupted images, few jumps are observed in the images which we proceed to obviate them, using codes of Image-translate.pro and Get-correct-offset.pro in IDL, and moreover, we write a code to compute offset of the images due to the remote locations of the loops (parts uniform in intensity and contains steady phenomena), and to apply its average to all of the images. The reason for choosing locations far away from the loops, is that the computed offset would be due to the images jumping, not of the likely inherent motion of the loops. Then, we use the MadMax or Octodirectional Maxxima of Convexities (OCM) filter (Koutchmy et al., 1989\cite{Koutchmy1989}) to reduce noises and improve raw images. This filter which was designed for noise reduction and resolution improvement of STEREO EUV images is also applicable for SDO images. This algorithm designed by Koutchmy (1989) is a special practical one in pattern recognition acting on the basis of using an operator. This operator selects the maximum of calculated convexities within eight directions around each pixel.
We write needful codes for cropping the images in desired regions, and then, we rotate and save the cropped images for all the time series. After studying the images, we select the subregion in which the semi-oscillatory phenomenon (in intensity) is observed. This event is visible in the images of the time duration between 14:05 and 14:20, on which we focus for more investigation. The loops of this subregion are extracted by using the method and related algorithms and codes explained below. Using these codes, the intensities along the loops are calculated. After testing the loops intensities of this subregion, we select the middle one (Fig.1.c) in which obvious intensity variation is observed. By comparison, we understand that the same phenomenon has significant differences in various wavelengths. For temperature extraction we consider the image in the middle of the abovementioned time interval, at 14:13:36.
For recognition of the loops, we use B-spline and improfile algorithms (by Bilinear interpolation method). We write a code in MATLAB using a 5-order polynomial B-spline function and the improfile algorithm. At the first step, the code asks us the desired number of the points we want to apply to extract the loop by them. However, by the power of improfile algorithm the intensity values between the points could be interpolated either. The improfile algorithm has three different methods to interpolate the points: Nearest-neighbor interpolation, Bilinear interpolation, and Bicubic interpolation. By examining these methods, we find that the Bilinear interpolation method is the best for our purpose and has less noise. In addition, by the help of Frenet-Serret algorithm, the perpendicular and parallel vectors could be also calculated in each point of the loop. To extract the loops, we use the Aschwanden method OCCULT (Oriented Coronal CUrved Loop Tracing, available in his homepage) either, to compare it and be more sure about our results. The extracted loop length is 95.3 in pixels which since every pixel in AIA is equal to $0.6$ arcsec, and each arcsec is $725km$ the loop length is equal to 41.4Mm.

\section{Background Subtraction and Intensity Profiles}

It is necessary to subtract the images background from the original images, specially for the processes, in which the intensity values are important, such as computing the thermal properties of the images or calculating the flux ratios of the loops. Subtraction of the background radiation is like applying a special filter. Due to the image specifications and intensity profiles, we should find the most appropriate method of background subtraction. The background in the EUV images includes any uniform intensity (either behind or front of the desired pattern) which does not depend on our desired underprocessing pattern and its intensity is annoying for the intensity of the understudy phenomenon (For more details about different methods of the background subtraction, see (Aschwanden et al., 2008)\cite{Aschwanden2008}.) However, uncertainty in the background subtraction, affects the related calculations and analysis. To subtract the background we write a code in which we use "strel" algorithm in MATLAB. This algorithm, which is derived from the work of Adams (1993) and works based on the characteristics of the selected part of the image, is useful \cite{Adams1993}. An example of the background subtraction of the image is observable in Fig.2. This figure shows 3D surface intensity profiles related to the selected image (at 14:13:36) at the wavelength of 171$A^0$, for the original image (above profiles), the original image after background subtraction (the middle profile), and the subtracted background (the lower one). The horizontal axis (ranges (0, 60) and (0, 100)) are the pixels coordinates (a part of Fig.1.c) and the vertical axis shows the intensity values.

%\begin{figure} %Fig.~2
%\includegraphics[width=0.9\columnwidth]{Figure2.pdf}
% %\centerline{\includegraphics[width=5cm]{Figure2.pdf}}
% \caption{The 3D surface intensity profiles related to the selected image (at 14:13:36) at the wavelength of 171$A^0$, for the original image (above profiles), the original image after background subtraction (the middle profile), and the subtracted background (the lower one). The horizontal axes are pixels coordinates and the vertical axis shows the intensity values.}
% \label{fig2}
%\end{figure}

\begin{figure} %Fig.~2
 \centerline{\includegraphics[width=0.9\textwidth]{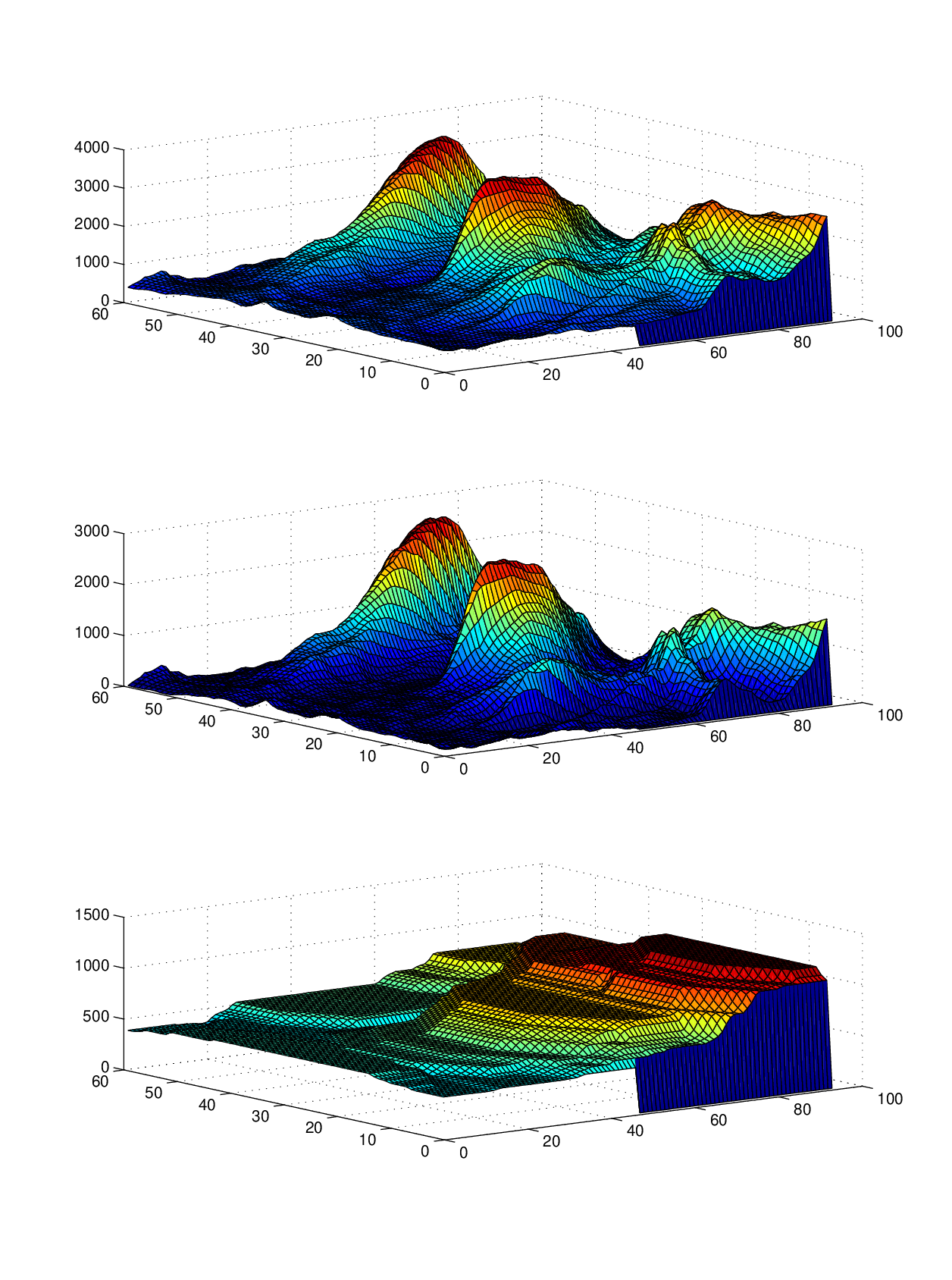}}
 \caption[]{The 3D surface intensity profiles related to the selected image (at 14:13:36) at the wavelength of 171$A^0$, for the original image (above profiles), the original image after background subtraction (the middle profile), and the subtracted background (the lower one). The horizontal axes are pixels coordinates and the vertical axis shows the intensity values.}
 \label{fig2}
\end{figure}

In figure 3, a sample of transversal intensity profiles have been plotted for one point in loop distance, as a representative of the loop, and for the wavelength of 171$A^0$ (for a part of region of Fig. 1.c.). In this figure, the intensity profiles could be seen in the direction perpendicular to the loop length (cross section of the loop), for the original intensity (in red), and the intensity profile after background subtraction (in blue). The transversal intensity profile is also plotted for the background (in pink). This shows how our background subtraction method works. After subtracting the background, we could plot the intensity profile along the loop.
%\begin{figure} %Fig.~3
%\includegraphics[width=0.9\columnwidth]{Figure3.pdf}
% %\centerline{\includegraphics[width=5cm]{Figure3.pdf}}
% \caption{The transversal intensity profiles, for a selected point of the loop in the direction perpendicular to the loop length (cross section of the loop), for the original intensity (in red), and the intensity profile after background subtraction (in blue), and the transversal intensity profile of the background (in pink). The horizontal axis shows the coordinates of the pixels along the cross section of the loops. And the vertical axis is the intensity values of the pixels.}
% \label{fig3}
%\end{figure}
\begin{figure} %Fig.~3
 \centerline{\includegraphics[width=0.9\textwidth]{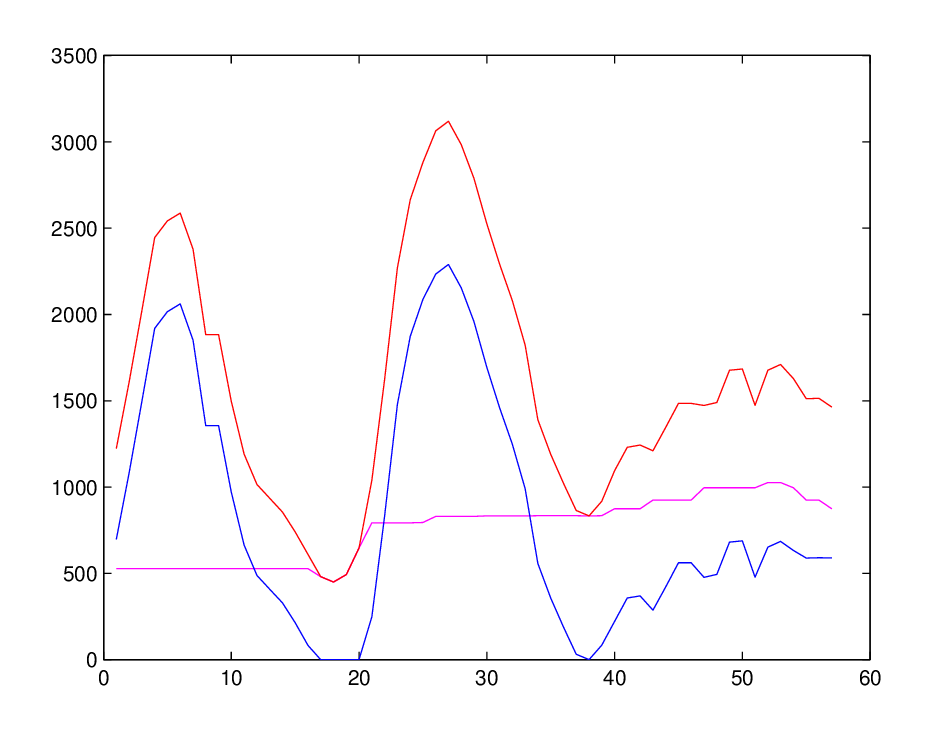}}
 \caption[]{The transversal intensity profiles, for a selected point of the loop in the direction perpendicular to the loop length (cross section of the loop), for the original intensity (in red), and the intensity profile after background subtraction (in blue), and the transversal intensity profile of the background (in pink). The horizontal axis shows the coordinates of the pixels along the cross section of the loops. And the vertical axis is the intensity values of the pixels.}
 \label{fig3}
\end{figure}
By pursuing the intensity profiles along the specified loop, the intensity event could be observed. Figure 4 shows the intensity profiles for a sample series of the selected loop of the foresaid region at the wavelength of 171$A^0$. Ten of intensity profiles are presented here, in the time duration of 13:57:12 till 18:00:00 with an equal time interval of 27min.  Background radiation in these images is subtracted by the method which was described above. As seen in this figure, the intensity varies along the loop in the time series of images. The intensity profiles along the loop show several increases (maximum) and decreases (minimum) in the observed period of time at the wavelength of 171$A^0$ (Fig. 4), and in the wavelength of 193$A^0$ the intensity variation is somewhat the same. The intensity variation at the wavelength of 171$A^0$ is more obvious while is also more or less obvious in the other two wavelengths.
While in the wavelength of 211$A^0$ this variation as a clear increase and decrease is less distinct.
%\begin{figure} %Fig.~4
%\includegraphics[width=0.9\columnwidth]{Figure4.pdf}
% %\centerline{\includegraphics[width=5cm]{Figure4.pdf}}
% \caption{Ten of intensity profiles for a sample series of the cropped images of the selected loop at the wavelength of 171$A^0$, in the time duration of 13:57:12 till 18:00:00 with an equal time interval of 27min between each two profiles. The horizontal axis is loop distance.  }
% \label{fig4}
%\end{figure}
\begin{figure} %Fig.~4
 \centerline{\includegraphics[width=0.9\textwidth]{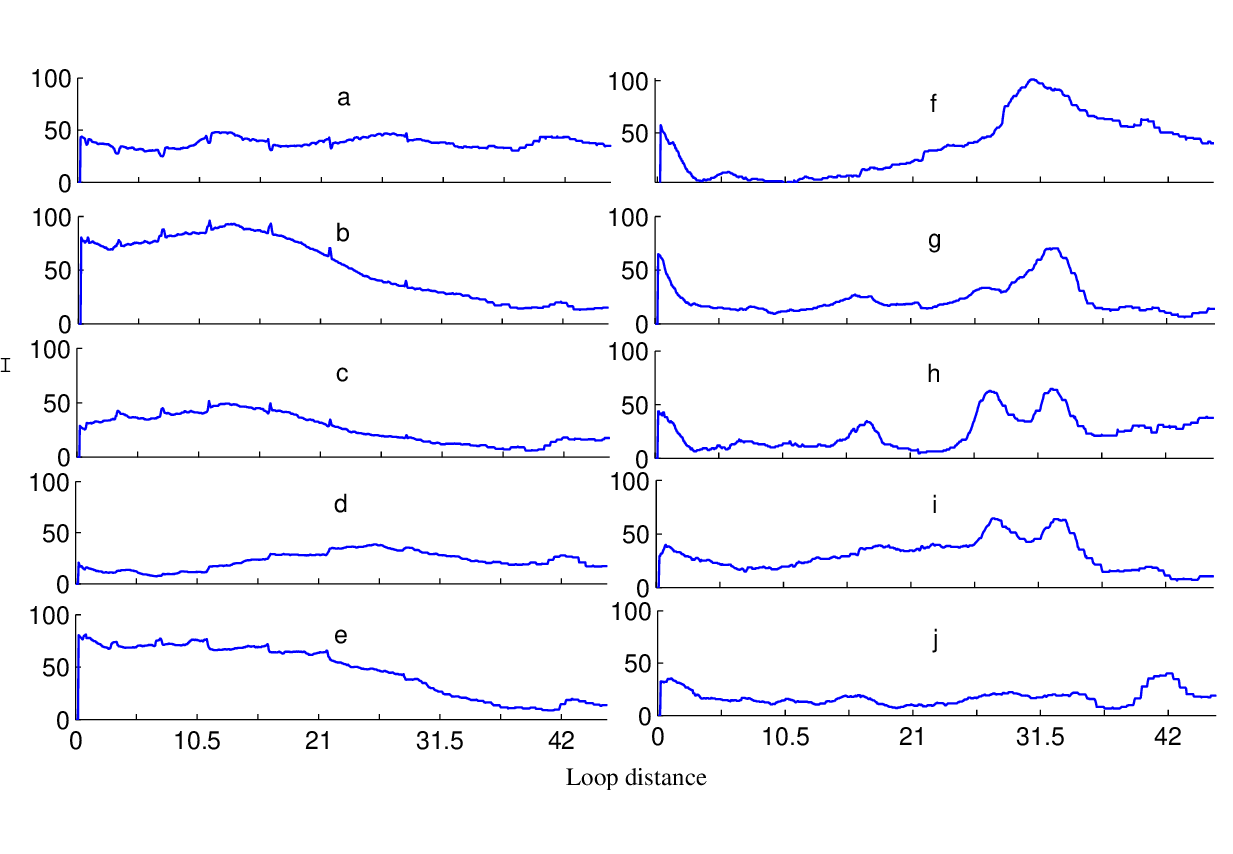}}
 \caption[]{Ten of intensity profiles for a sample series of the cropped images of the selected loop at the wavelength of 171$A^0$, in the time duration of 13:57:12 till 18:00:00 with an equal time interval of 27min between each two profiles. The horizontal axis is loop distance.}
 \label{fig4}
\end{figure}
\section{Thermal Investigation}

A spectral line forms at specific temperature. Hence, broadband filters are responsible for a temperature response function resulting from the observed wavelength range. Different response functions for various wavelengths of SDO instruments exist in IDL software under SSW (Solar Soft Ware). Most of the temperature profiles indicate that an AIA image does not correspond just to a unique temperature, but often to two temperature peaks and a broad range of temperature. To interpret the AIA images, it is important to understand that at which temperature the plasma is. Figure 5 (left column) shows the temperature response functions for different filters versus the logarithm of the temperature (Brown et al., 2010)\cite{Brown2010}.
To allow comparison of temperature response functions of different filters and loops intensity at different wavelengths, we need to have dimensionless functions. Therefore, we use temperature ratios and intensity ratios. Figure 5 (right column) shows the ratios of temperature response functions. Fig. 6. (right) shows the ratios of the intensities at different wavelengths along the points of the selected loop, for different wavelengths of 171$A^0$ to 193$A^0$ (upper one), 211$A^0$ to 193$A^0$ (middle one), and 211$A^0$ to 171$A^0$ (lower one), versus the loop distance. These two profiles set (Fig.5 (right) and Fig.6 (right)) are both dimensionless and could be used for thermal analysis of the desired loop.
At the wavelength of 171$A^0$, we observe a region with high intensity which is clearly identified in the movie of this region as well. Since this intense region can not be seen at the wavelength of 193A0, it seems that cooling of the mentioned loop happens in this region. For this loop, the highest intensity ratio of 171$A^0$ to 193$A^0$ is about $0.76$ (Fig. 6) and this digit in Fig. 5 (right column) corresponds to the temperature of $1.3$ million degrees. In this calculation, the upper value of the temperature, in the profile of temperature response ratios, is considered; since this is the expected order of value for the temperature of the loops. This temperature value is around the temperature peak of the mentioned loop's internal. Due to the Fig.6 (right), the highest values of the intensity ratios at the wavelengths of 211$A^0$ to 193$A^0$ (the middle one), and 211$A^0$ to 171$A^0$ (the lower one) are $0.22$ and $0.25$ and based on the Fig.5 (the right column) these digits corresponds to temperature values around $10M$ and $1.4M$ Kelvin in sequence, which are related to the hotter and more superficial points of this loop.

%\begin{figure} %Fig.~5
%\includegraphics[width=0.9\columnwidth]{Figure5.pdf}
% %\centerline{\includegraphics[width=5cm]{Figure5.pdf}}
% \caption{Left column: the temperature response functions for different filters of 171$A^0$ (upper one), 193$A^0$ (middle one), and 211$A^0$ (lower one), versus the logarithm of the temperature \cite{Brown2010}. Right column: The ratios of the temperature response functions for different filters of 171$A^0$ to 193$A^0$ (upper one), 211$A^0$ to 193$A^0$ (middle one), and 211$A^0$ to 171$A^0$ (lower one), versus the logarithm of the temperature. }
% \label{fig5}
%\end{figure}
%
%\begin{figure} %Fig.~6
%\includegraphics[width=0.9\columnwidth]{Figure6.pdf}
% %\centerline{\includegraphics[width=5cm]{Figure6.pdf}}
% \caption{The intensity profiles along the extracted loop in different wavelengths versus the loop distance. Right: The ratios of the intensities at different wavelengths along the specified loop, for different wavelengths of 171$A^0$ to 193$A^0$ (upper one), 211$A^0$ to 193$A^0$ (middle one), and 211$A^0$ to 171$A^0$ (lower one), versus the loop distance.}
% \label{fig6}
%\end{figure}
\begin{figure} %Fig.~5
 \centerline{\includegraphics[width=0.9\textwidth]{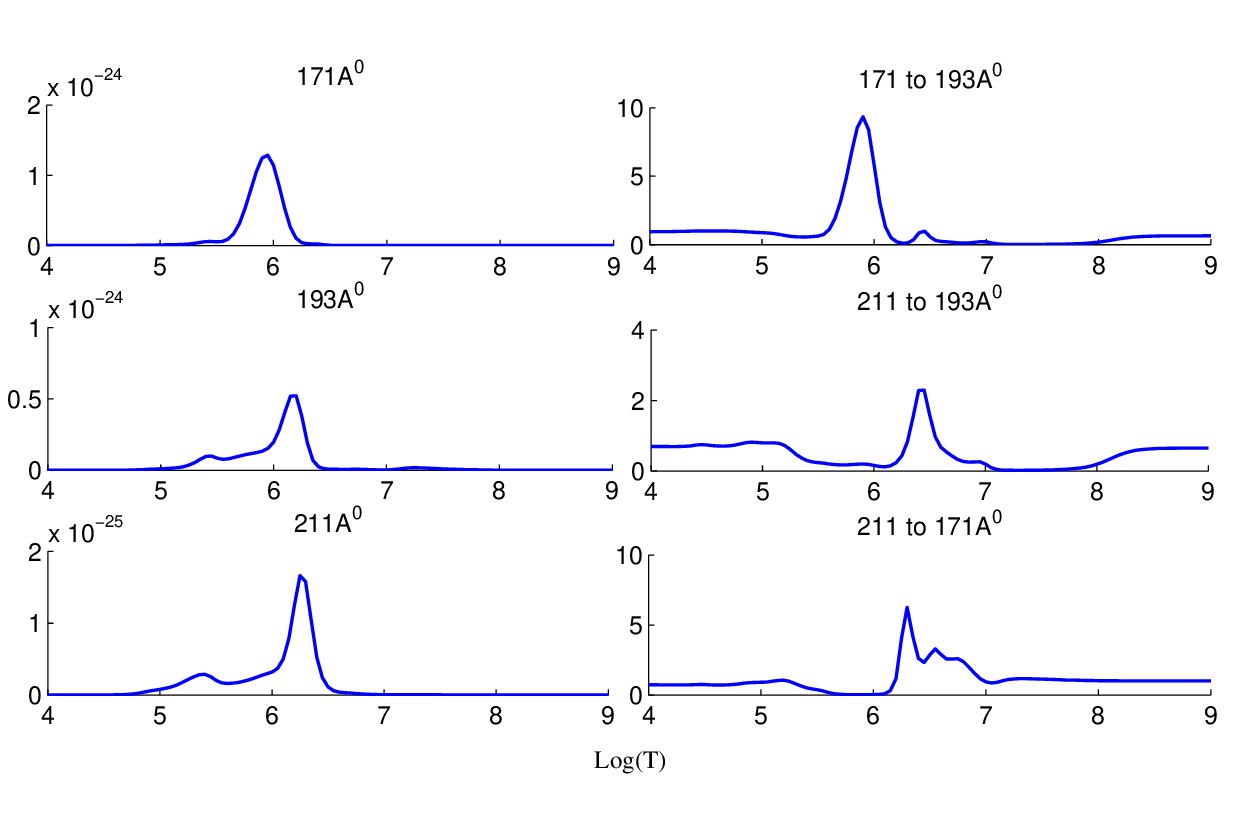}}
 \caption[]{Left column: the temperature response functions for different filters of 171$A^0$ (upper one), 193$A^0$ (middle one), and 211$A^0$ (lower one), versus the logarithm of the temperature \cite{Brown2010}. Right column: The ratios of the temperature response functions for different filters of 171$A^0$ to 193$A^0$ (upper one), 211$A^0$ to 193$A^0$ (middle one), and 211$A^0$ to 171$A^0$ (lower one), versus the logarithm of the temperature.}
 \label{fig5}
\end{figure}
\begin{figure} %Fig.~6
 \centerline{\includegraphics[width=0.9\textwidth]{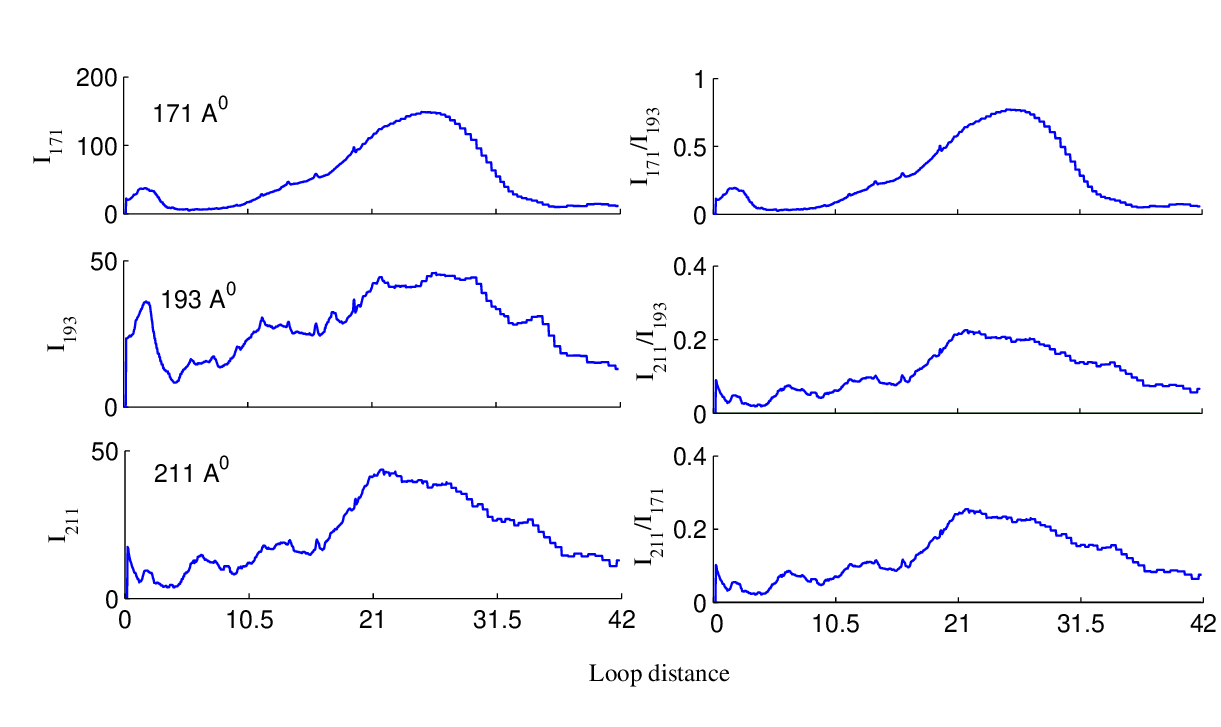}}
 \caption[]{The intensity profiles along the extracted loop in different wavelengths versus the loop distance. Right: The ratios of the intensities at different wavelengths along the specified loop, for different wavelengths of 171$A^0$ to 193$A^0$ (upper one), 211$A^0$ to 193$A^0$ (middle one), and 211$A^0$ to 171$A^0$ (lower one), versus the loop distance.}
 \label{fig6}
\end{figure}
\section{Results and Conclusion}

In this article, we explained how the active region of 11092 was selected, for thermal investigation of solar active region. The data includes the images recorded by the AIA instrument of the SDO satellite at different wavelengths. We explained that after observing the time-series images, selecting the desired thermal event, modifying image's jumping and the effect of the sun rotation, the desired subregion was cut from all of the time series images. Then, by subtracting the background intensity and extracting the loops, we plotted the intensity profiles.
As has been pointed above, by comparing images in different wavelengths, we realized that the phenomenon of the same region has significant differences in various wavelengths. We did this comparison in the different wavelengths of 171$A^0$, 193$A^0$, and 211$A^0$, which are related to different states of ionized iron. As we seen, in the results of comparing the images in different wavelengths, the loop lines are more obvious at the cooler wavelengths. For instance, at the wavelength of 171$A^0$, the loop lines are more central and sharper. While at hotter wavelengths of 193A0, the loops are more scattered and dispersed. Paying attention at loop images in different wavelengths, it is clear that the loops lines are more coherent and specific, in sequence at the wavelength of 211$A^0$ relative to 193A0, and at the wavelength of 193$A^0$ relative to 171A0.These images can be looked as thermal maps. Cooler wavelengths, indeed show the inner and more central information of the loop, while hotter wavelengths, which are more scattered, indicate the information of the surface of the loop which (the surface) is in higher temperature.
By pursuing the intensity profiles along the specified loop in the considered time interval, the intensity event could be observed. The intensity variation at the wavelength of 171 A0 is more obvious than the other two wavelengths. In this loop, the intensity profiles show several obvious increases (maximum) and decreases (minimum) in the observed period time at the wavelength of 171 $A^0$, and in the wavelength of 193 $A^0$, the intensity variation is somewhat the same. While in the wavelength of 211 $A^0$, this variation as a clear increase and decrease is less distinct.
Afterwards, by the help of the information which various filters of the AIA instrument provided us, we achieved thermal response ratios and intensity ratios for the active region of 11092 at different wavelengths. By using this information, we did thermal investigation in desired event. In the selected images at the wavelength of 171A0, we see an area with high intensity which does not observe at the wavelength of 193$A^0$. In this subregion in which the cooling of the mentioned loop occurs, the upper limit of the intensity at 171$A^0$ to 193$A^0$ was achieved to be about $0.76$ (upper in the Fig. 5, right) and this digit (due to the Fig.4, right, upper one) corresponds to the temperature of $1.3$ Million degree of Kelvin, which is about the maximum internal temperature of this loop. Also, the highest values of the intensity ratios at the wavelengths of 211$A^0$ to 193A0 (the middle one), and 211$A^0$ to 171$A^0$ (the lower one) are $0.22$ and $0.25$, which corresponds to temperature values around $10M$ and $1.4M$ Kelvin, in sequence, which are related to the hotter and more superficial points of this loop. However it is difficult to determine the exact temperature value. One reason of this inaccuracy is the high inclination in this zone of the temperature response ratios profile in Fig. 5 (the right column), and it means that a small change in intensity ratios, causes a wide change in temperature. The other reason is uncertainty in the background subtraction of image, and since the amount of the background radiation is very important in the value of the achieved temperature, this uncertainty is inevitable.
Aschwanden et al. 2015 \cite{AschwandenB2015} developed a method for extracting the loop temperature which is based on Gaussian fit for Differential Emission Measure, named spatially-synthesized Gaussian DEM forward-fitting method. Here, we developed a different method using the intensity ratios of the images in three different wavelengths and the temperature response functions of AIA to extract the temperature of the loop. While DEM method uses 6 wavelengths to calculate the temperature, we use 3 ones and DEM evaluates one temperature for each cross point of the loop fitting Gaussian profile while our method estimates different temperature in cross point which could be a representative of the temperature of different parts of the loop (surface or inner layer).  Beside that DEM method could also calculate the errors so easily and its code is available online, so it is more accurate and accessible. As we compared their method, the temperature which DEM estimates for this loop varies from $5$ to $7.3$Mk depending on different points along the loop. This temperature range is exactly the temperature range which Aschwanden et al. 2015 proposed and expected for a loop to have (this range is fed from the beginning as a limitation to the numerical program). The temperature which DEM achieves for this loop differs for different parts of the loop and covers the initial range. Therefore our estimation of the temperature is totally at the same order of magnitude but the minimum estimated temperature is lower in our work. The code we developed in this work can be useful in analyzing temperature of the coronal loops using 3 wavelengths of 171$A^0$, 193$A^0$ and 211$A^0$.

\section*{Acknowledgment}
This project has been done under the support and collaboration of the Max-Planck Institut für Sonnensystemforschung (MPS) and I acknowledge the warm hospitality during my research visit to the  Max-Planck Institut für Sonnensystemforschung (MPS). I would like to sincerely acknowledge Prof.Bernd Inhester, Dr. Davina Innes, Prof.Sami Solanki and Prof.Hossein Safari. I also appreciate helpful and constructive comments from an anonymous referee.

\end{document}